# Slip intermittency and dwell fatigue in titanium alloys: a discrete dislocation plasticity analysis

Yilun Xu[1,2]*, Felicity Worsnop[1,3], David Dye[1] and Fionn P.E. Dunne[1]

[1]Department of Materials, Imperial College, London SW7 2AZ, UK

[2]Institute of High Performance Computing, Agency for Science, Technology, and Research (A∗STAR), Singapore 138632, Singapore

[3]Department of Materials Science and Engineering, Massachusetts Institute of Technology, Cambridge, MA, 02139, USA

**Abstract**

Slip intermittency and stress oscillations in titanium alloy Ti–7Al–O that were observed using *in-situ* far-field high energy X-ray diffraction microscopy (ff-HEDM) are investigated using a discrete dislocation plasticity (DDP) model. The mechanistic foundation of slip intermittency and stress oscillations are shown to be dislocation escape from obstacles during stress holds, governed by a thermal activation constitutive law. The stress drop events due to <a>-basal slip are larger in magnitude than those along <a>-prism , which is a consequence of their differing rate sensitivities, previously found from micropillar testing. It is suggested that interstitial oxygen suppresses stress oscillations by inhibiting the thermal activation process. Understanding of these mechanisms is of benefit to the design and safety assessment of jet engine titanium alloys subjected to dwell fatigue.

---

[1] Corresponding author: yilun.xu@imperial.ac.uk;xu_yilun@ihpc.a-star.edu.sg

# 1 Introduction

Cold dwell fatigue (CDF) has been extensively recognised to be a key life-limiting failure mode for large jet engine compressor and fan discs (Pilchak and Gram, 2022), and an issue which is of significant safety concern. In dwell, fatigue lifetime is significantly reduced (Bache, 2003) compared to conventional low cycle fatigue (LCF) when titanium alloys are subjected to a stress hold at peak load and in the vicinity of room temperature (termed low cycle dwell fatigue, LCDF). Facet formation along the <a>-basal slip plane in the α phase (HCP crystal structure) of titanium alloys has been considered to be the micro-scale driver for cold dwell fatigue (Pilchak and Williams, 2010), and hence it is crucial to investigate the micro-scale stresses (Sinha et al., 2006) within titanium alloys that initiate the formation of such facetted features.

Dislocation motion has long been observed to be intermittent, using *in situ* straining in the TEM (Ispanovity et al., 2014), micropillar loading experiments (Hu et al., 2018), creep tests (Miguel et al., 2001) and more recently using *in-situ* far-field high energy X-ray diffraction microscopy (ff-HEDM) (Beaudoin et al., 2017), with the overall collective motion termed dislocation avalanches (Dimiduk et al., 2006). This gives rise to dynamic variations in local stress state, which may then impact upon the process of fatigue crack initiation, e.g. in an adjacent hard grain. Recently, we made observations on the effect of O content in Ti–7Al–O alloys on the extent of such stress oscillations using ff-HEDM. Stress fluctuation events during the hold period were shown to be dependent both on the slip system type (*i.e.* basal, prismatic or pyramidal) and the interstitial oxygen content (Worsnop et al., 2022). Stress oscillations during the stress hold period exacerbate load shedding (Venkatramani et al., 2007) from soft to hard grains, raise the basal stress in the hard grain and eventually lead to an early facet formation. Therefore it is crucial to reveal the mechanisms of the slip intermittency and stress drops at the microstructure scale, in order to identify the factors governing the stress drop events and provide insight into the optimisation of alloying and microstructure in titanium alloys.

Slip intermittency has been extensively observed during micro-scale compression testing, manifesting itself as instantaneous stress oscillations in the stress-strain response (Dimiduk et al., 2006; Hu et al., 2018; Ispanovity et al., 2022; Uchic et al., 2004). The stress oscillations have been attributed to sudden and inhomogeneous dislocation bursts as a random number of dislocations emit from nucleation sources subjected to load increments (Michael and Aifantis, 2006; Rizzardi et al., 2022a; Rizzardi et al., 2022b; Weiss et al., 2019; Zaiser and Moretti, 2005; Zaiser and Nikitas, 2007). Many numerical efforts have been made to provide mechanistic understanding at the dislocation or sub-dislocation scale for this slip intermittency, such as those using three-dimensional discrete dislocation plasticity (Crosby et al., 2015; Kurunczi-Papp and Laurson, 2021; Kurunczi-Papp and Laurson, 2023; Weiss and Marsan, 2003) and atomistic-scale approaches (Cao et al., 2018). However, previous efforts focused on the stress oscillation during continuous loading as opposed to those under a stress hold, which is more relevant to the situation in load controlled fatigue, such as that occurring in titanium jet engine components subjected to cold dwell fatigue. In addition, the high computational resource demand of those approaches inhibits the collection of meaningful statistics. Therefore, a more specific modelling framework is required to investigate local stress drops at the slip scale and their influence on the dwell fatigue performance of Ti alloys in a high-throughput way.

A two-dimensional discrete dislocation plasticity (DDP) methodology is utilised to describe the activities of individual dislocations within grains, and hence has been applied to understand the localised micro-deformation mechanisms (Tarleton et al., 2015) that occur due to collective dislocation motion and are affected by microstructure e.g. (Bergsmo et al., 2022; Xu et al., 2019). By incorporating both rate- and temperature- sensitivity which result from thermally activated dislocation escape from obstacles, a temperature-enhanced DDP framework has been utilised to provide mechanistic understanding of the thermomechanical alleviation (TMA) effect in Ti alloys, successfully predicting fatigue life under variations in temperature and stress loading conditions (Xu et al., 2020a). The intrinsic random and discrete nature of the DDP methodology (Deshpande et al., 2001; Xu et al., 2021) is able to reflect the stress oscillations that have been extensively observed in the material response subject to micro-pillar compression (Dimiduk et al., 2006; Uchic et al., 2009). The relatively high computational efficiency of 2D DD facilitates statistical analysis, such as the generation of Complementary Cumulative Distribution Functions, CCDFs (Zhang et al., 2017).

In the following sections of this paper, we begin with a brief summary of the key findings of our previous experimental *in situ* ff-HEDM testing on Ti–7Al–O (Section 2). The DDP framework, model and the material properties are reported in Section 3. The DDP simulations results are detailed in Section 4 with discussion and comparison to experiments, which are followed by some concluding remarks.

## 2 Slip intermittency and stress drops analysed in the HEDM experiments on Ti–7Al

High energy X-ray diffraction microscopy (HEDM) analysis was performed on polycrystal samples of Ti–7Al–O alloys, and stress bursts during creep were measured at the scale of individual grains *in situ* at room temperature. The full account of the experimental findings is given in (Worsnop et al., 2022), and hence only the key information is summarised in Figure 1. Specimens were heat treated to generate equiaxed grains with average grain size about 50 µm, and the c-axes of the grains were found to deviate away from the loading direction such that the majority of the grains were in a soft crystallographic orientation favouring <a> slip when loaded along the bar (rolling) direction, Figure 1 (a). *In situ* creep tests were performed with far-field HEDM, enabling time-resolved measurement of grain-average elastic strain states, under an applied stress of 80% of the yield stress. From one scan to the next, the local stress in each grain may either rise or fall, Figure 1 (c); drops in normalised resolved shear stress, $\Delta \bar{\tau} = \Delta(\tau/\tau_c)$, where τ and $\tau_c$ are the measured shear stress and critical resolved shear stress, respectively, were used to quantify slip avalanche events. Grains with stress states favouring basal <a> slip were found statistically to show larger stress drops compared to prism <a>, Figure 1 (d), while an increased interstitial oxygen concentration was found to smooth out the slip intermittency and stress drops, Figure 1 (e).

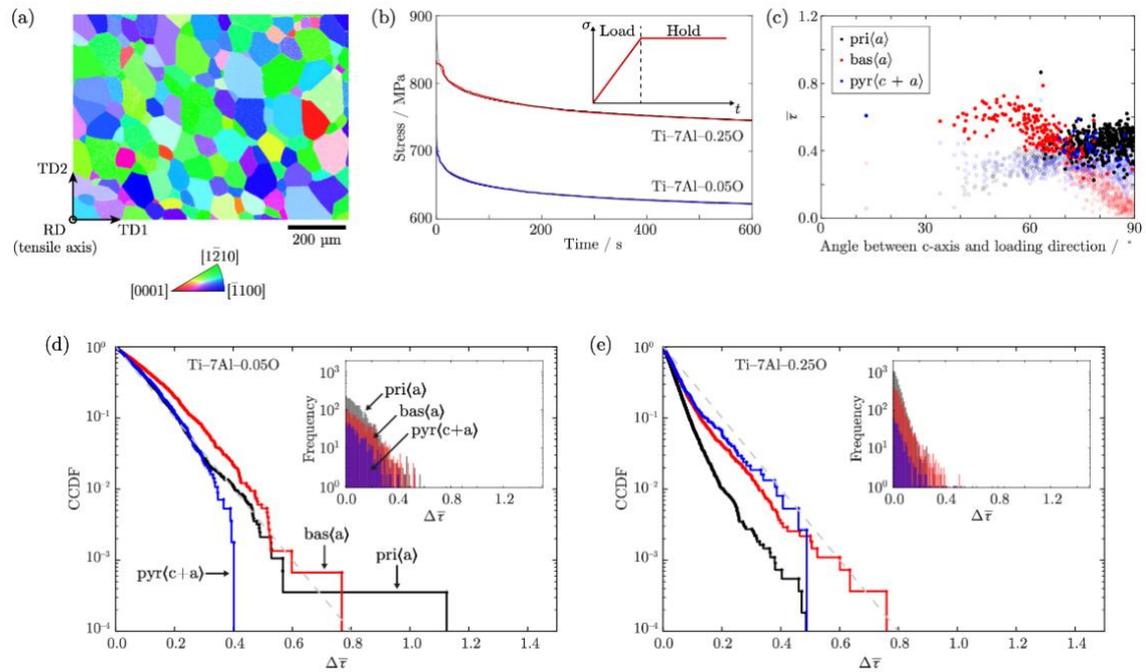

Figure 1. The material and summary of experimental results. (a) An inverse pole figure (IPF) electron backscatter diffraction map showing the rolling (micro)texture of the Ti–7Al–O alloys with respect to the rolling direction, which was also the loading direction in testing. Note that there are few grains with their c-axis oriented close to the loading direction, suggesting a limited number of 'hard' grain orientations in the HEDM analysis volume. (b) Stress relaxation curves for low and high O alloys during a strain-controlled hold after deformation to ~1.5% applied strain. Inset is a schematic of the macroscopic loading conditions. (c) For a ff-HEDM dataset measured at a macroscopically elastic state, normalised resolved shear stresses $\bar{\tau} = \tau/\tau_c$, are shown for each slip system in each measured grain. Three points are shown for each grain: Filled circles show the most highly stressed slip system for a grain, i.e. the system most likely to initiate slip, while open circles show the others, showing the preference for pri<a> slip at high c-axis declinations and for bas<a> at low declinations. (d) and (e) Cumulative complementary distribution functions (CCDFs) and histograms of the stress drops, $\Delta\bar{\tau}$, for low- and high-oxygen alloys, during creep testing at constant load. The dashed lines are a visual guide only. The slip events in <a>-basal slip systems are larger in magnitude than those in <a>-prism in both cases, and a raised interstitial O content considerably refines the stress drop events, smoothing out slip intermittency.

The dislocation structures produced after stress relaxation were also observed, Figure 2; these observations are further detailed in the experimental paper (Worsnop et al., 2022). Dislocation densities of approximately 4.8×10$^{13}$ m$^{-2}$ in Ti–7Al–0.05O compared to 1.2×10$^{14}$ m$^{-2}$ in Ti–7Al–0.25O (assuming foil thickness ~100 nm) were estimated from these representative images, i.e. approximately 2.4 times higher in the high-O alloy under these stress relaxation conditions. For comparison of obstacle density between the experiment and the model, we may estimate the spacing of oxygen interstitials in each of the real alloys. Assuming a random, homogeneous spatial distribution of the oxygen atoms in the Ti–7Al material, the O atoms in the Ti–7Al–0.05O alloy had an average spacing of 21 nm, and in the Ti–7Al–0.25O alloy, an average spacing of 13 nm. This corresponds to a roughly 2.5 times higher two-dimensional O atom density in the high-O alloy than in the low-O alloy. Considering the STEM observations, the raised O content causes a 2.4 times increase in the observed dislocation density, in qualitative agreement with a doubling in dislocation density upon doubling the obstacle density in DDP modelling (see later, Figure 9). The TEM observations therefore illustrate the potential role of interstitial O in the increasing dislocation density in strained Ti–Al–O alloys, which could be of importance in titanium alloy performance optimisation.

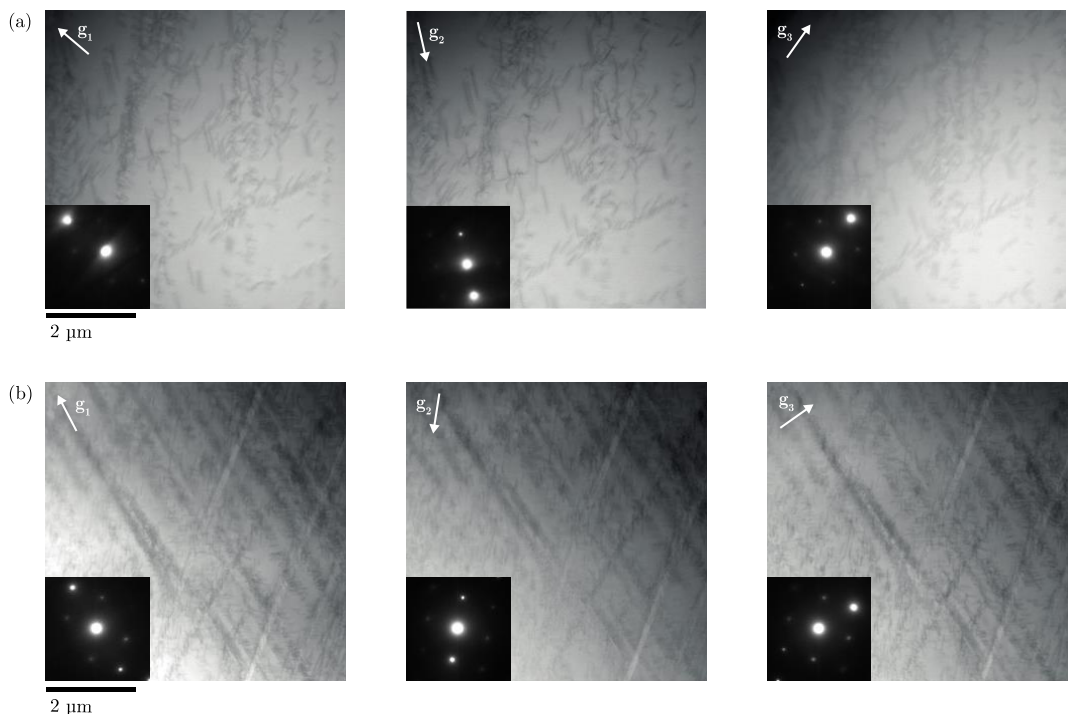

Figure 2. Dislocation structures produced after stress relaxation testing in (a) Ti–7Al–0.05O and (b) Ti–7Al–0.25O (wt.%). In this test, samples were loaded to ∼1.5% macroscopic strain and held in strain control for 10 minutes while stress relaxation took place. Images were obtained in STEM mode using an [01$\bar{1}$1] beam direction, with a viewing direction approximately parallel to the loading direction and two-beam conditions used to excite $g_1 = [2\bar{1}\bar{1}0]$, $g_2 = [\bar{1}10\bar{1}]$, $g_3 = [0\bar{1}12]$.

## 3  Methodology

This section begins with a brief introduction of the discrete dislocation plasticity (DDP) formulations adopted to model the micro-scale behaviours of Ti–7Al–O alloys. The DDP materials properties for Ti–7Al–O alloys, which are obtained from a calibration to experiment, are then shown. The DDP model that is used to investigate the slip intermittency and stress drops is described at the end of the section.

### 3.1  Discrete dislocation plasticity formulations

A two-dimensional and small strain discrete dislocation plasticity framework is adopted to investigate the micromechanical behaviour of the Ti–7Al–O alloys. The superposition scheme (Van der Giessen and Needleman, 1995) and the computational framework by (Tarleton et al., 2015) are employed to solve the discrete dislocation fields and the boundary value problem. The full set of DDP constitutive laws have been detailed previously, and hence they are only summarised here.

The total displacement field $\boldsymbol{u}$ is decomposed into a $\tilde{\boldsymbol{u}}$ field that sums the displacements of discrete dislocation contributions, and a correction field $\hat{\boldsymbol{u}}$ that corrects for the imposed boundary conditions. The complete solution of stress $\boldsymbol{\sigma}$ and strain $\boldsymbol{\epsilon}$ fields are obtained by linear superposition of the corrected image field ($\hat{\ }$) and the sum of individual dislocation fields ($\tilde{\ }$), where $n$ is the total number of dislocations at the current time increment:

$$\boldsymbol{u} = \sum_{i=1}^{n} \tilde{\boldsymbol{u}}_i + \hat{\boldsymbol{u}}$$

$$\sigma = \sum_{i=1}^{n} \tilde{\sigma}_i + \hat{\sigma} \quad (1)$$

$$\epsilon = \sum_{i=1}^{n} \tilde{\epsilon}_i + \hat{\epsilon}$$

Dislocations are pinned at obstacles, representing imperfections such as interstitial atoms within crystals, when gliding along slip systems. A pinned dislocation is released when its residence time reaches a critical value $t_{obs}$ that is governed by thermal activation (Zheng et al., 2016c). The critical pinning time $t_{obs}$ is:

$$\frac{1}{t_{obs}} = \Gamma = \frac{v_D b}{l_{obs}} \exp\left(-\frac{\Delta H}{k\theta}\right) \sinh\left(\frac{\tau \Delta V}{k\theta}\right) \quad (2)$$

where $v_D$ is the attack frequency of dislocations to jump obstacles, $l_{obs}$ the obstacle spacing (hence determined by the obstacle density), $\Delta H$ the activation energy, $\Delta V$ the corresponding activation volume, $k$ the Boltzmann constant, $b$ the Burgers vector, $\tau$ the resolved shear stress on a given system and $\theta$ is temperature. The residence time is determined from both the material properties, i.e. activation energy and activation volume, which differ for HCP slip systems in Ti alloys (Xiong et al., 2020), and the local shear stress status. The residence time of a dislocation pinned at obstacles increases until the critical pinning time is achieved (i.e. the dislocation is released). The introduction of the pinning time provides an intrinsic time-dependent (*i.e.* strain rate-sensitive) material behaviour, which manifests itself as stress relaxation under constant strain control or creep under constant stress control. Zheng et al. (Zheng et al., 2016c) has shown that the rate-sensitivity of Ti alloys is predominately due to the thermally activated dislocation escape (eq.(2)) from obstacles, rather than due to contributions from dislocation nucleation and mobility (Fan et al., 2021). For the relatively low strain rates considered in the present study, high-rate effects such as elastic dynamic stresses and stress waves (Gurrutxaga-Lerma et al., 2017)) are ignored.

In eq.(2), the time constant of the dislocation escape events is governed by the two independent material properties, $\Delta H$ and $\Delta V$, as well as the obstacle density $\rho_{obs}$ at a given temperature, where the latter is given by the average obstacle distance $l_{obs}$:

$$\rho_{obs} = 1/\sqrt{l_{obs}} \tag{3}$$

Smaller-scale simulations, such as those using density-function-theory (Yu et al., 2015), *ab initio* (Chaari et al., 2019; Rodney et al., 2017) and first-principle (Ghazisaeidi and Trinkle, 2014), have suggested that the oxygen interstitials and atoms could impede dislocation motion along slip systems and hence the dislocation dynamics are impacted by the oxygen content within Ti alloys.

The requirement of plane strain is achieved by confining the HCP slip systems to three $\langle a \rangle$-prismatic/basal slip systems in 'soft' oriented grains and two 1st order $\langle c + a \rangle$-pyramidal plus one $\langle a \rangle$-basal slip systems in 'hard' oriented grains, with respect to the loading conditions. The assumption is justified by the experimental observation of the planar dislocation structures of Ti alloys during dwell fatigue tests (Joseph et al., 2018; Suri et al., 1999). The events of dislocation climb are not considered in the DDP modelling as the samples were tested at a relatively low temperature.

## 3.2 Material properties

A calibration process is carried out for extracting the DDP material properties for aged Ti–7Al–O with low oxygen content by matching the predicted macroscopic stress response to the experimental response measured in uniaxial tension testing, Figure 3(a), and under stress relaxation, Figure 3(b) at room temperature. A polycrystal DDP model (results depicted in the inset contour plots) that has been established (Xu et al., 2020a) for similar α-titanium alloys such as Ti-834, Ti-6242 and Ti6246 is used here to reflect the microstructural feature of the soft-hard grain combination in the Ti–7Al–O alloy. The consistency between the calculated and experimentally measured stress response indicates that the calibrated DDP model is able to reflect the mechanical behaviour of Ti–7Al–O. It is noted that the loading rates imposed are different for the uniaxial tension and stress relaxation test, and hence the calibrated thermal activation governing parameters are able to capture the time-dependent feature (i.e. creep), which is determined by the dislocation escape events from obstacles (Xu et al., 2020b). The calibrated DDP parameters, which include source and obstacle density $\rho_{nuc}$, $\rho_{obs}$ nucleation strength of dislocation source $\bar{\tau}_{nuc}^{<a>}$, activation energy $\Delta H$ and volume $\Delta V$, are tabulated in Table 1.

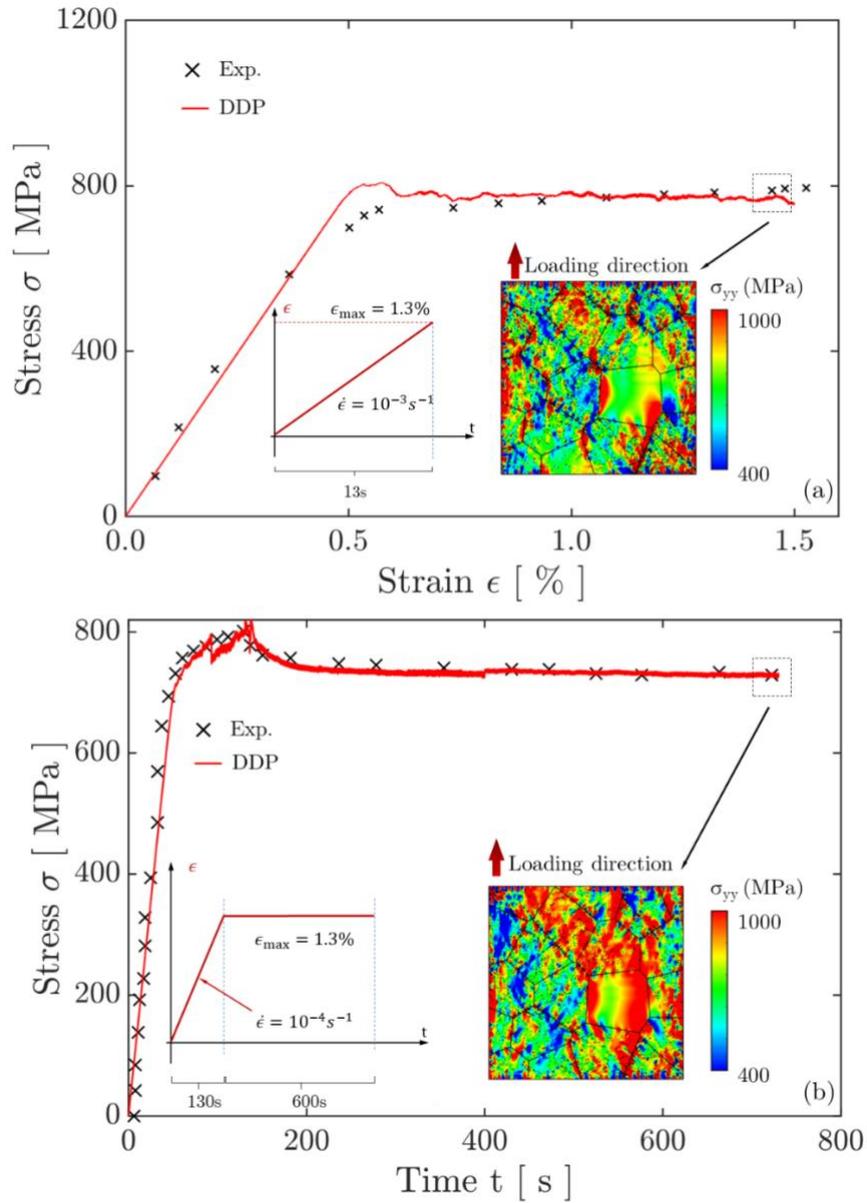

Figure 3. The calibration for obtaining the DDP material properties of the polycrystal Ti–7Al–O alloy (with low oxygen content and aged) to the tests of (a) the uniaxial tension with applied strain rate $\dot{\epsilon} = 10^{-3}\,s^{-1}$ and (b) the stress relaxation test with applied strain rate $\dot{\epsilon} = 10^{-4}\,s^{-1}$. The inset contour plots show the stress field predicted using DDP method at the end of the corresponding load history.

Table 1. The DDP material properties for Ti–7Al–O alloy (aged and with low oxygen content) that are obtained from calibration to the tension and stress-relaxation tests.

| $G$(GPa) | $\nu$ | $b$(nm) | $\rho_{nuc}$(μm$^{-2}$) | $\rho_{obs}$(μm$^{-2}$) | $\bar{\tau}_{nuc}^{<a>}$(MPa) |
|---|---|---|---|---|---|
| 33.0 | 0.45 | 0.295 | 30 | 200 | 450 |
| $\nu_D$(Hz) | $B$(Pa·s) | $\eta$(Pa·s) | $k$(JK$^{-1}$) | $\Delta H$(J/atom) | $\Delta V$ |
| $10^{11}$ | $0.5 \times 10^{-4}$ | $90B$ | $1.38 \times 10^{-23}$ | $9.20 \times 10^{-20}$ | $0.50b^3$ |

### 3.3 DDP model for investigating slip intermittency in Ti–7Al–O

A Ti–7Al–O single crystal with a 20 μm grain size was modelled using the two-dimensional discrete dislocation plasticity (DDP) methodology, with the crystal embedded in a bounding elastic region to avoid boundary effects. The crystal was uniaxially loaded along the y-direction, with the left and bottom boundaries constrained accordingly, Figure 4(a). There are three <a>-prism or <a>-basal slip systems allocated to the DDP window to satisfy the plane strain condition (Van der Giessen and Needleman, 1995), and nucleation sources and obstacles are randomly populated along the slip systems with a specified density as sketched in Figure 4 (b). The single crystal c-axis inclination was rotated (the red arrow) in order to vary the orientation of the three slip systems with respect to the loading direction. The loading history that aims to replicate the creep experiments is shown in Figure 4 (c), and the holding span represents a typical stress hold period in dwell fatigue test (Bache, 2003). The stress is averaged within the sufficiently large centre square (18 μm × 18 μm depicted in the framed zone in Figure 4 (d)) of the single crystal to avoid stress concentrations due to dislocations piling up near the grain boundary.

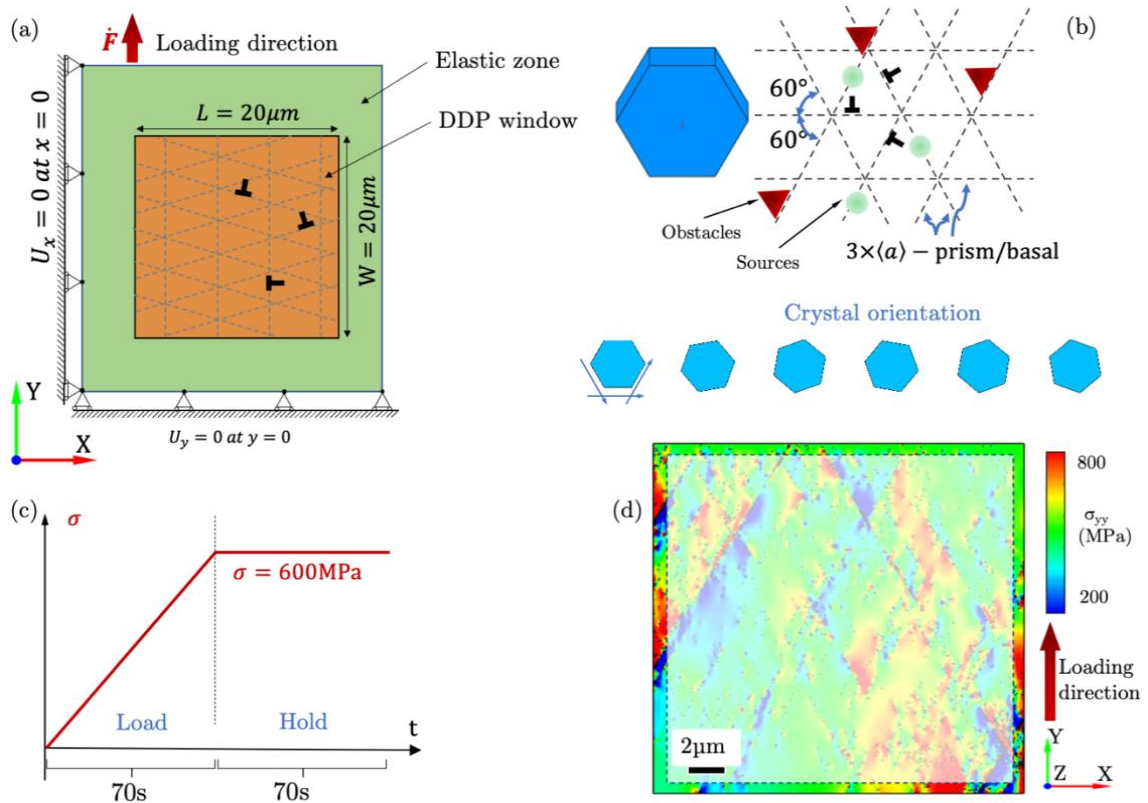

Figure 4. Discrete dislocation plasticity (DDP) modelling of a single Ti–7Al–O crystal under stress holding. (a) The single crystal surrounded by an elastic region to avoid boundary effects. (b) Schematic configuration of the dislocations gliding along three <a>-basal or prism slip systems, which are also populated with dislocation nucleation sources and obstacles. The slip system can be rotated following the crystallographic orientation of the single crystal. (c) The loading history applied on the single crystal that includes one stress loading and one hold period. (d) A typical stress field (the component along the loading direction) of the single crystal during the stress hold. The stress sampling region is allocated within the framed region shown, the boundaries of which are set 0.5 µm away from the single crystal boundary to avoid stress concentrations.

## 4 Results

A typical stress analysis for the stress drops within a Ti–7Al–O single crystal using the discrete dislocation plasticity method is shown in Figure 5. For one instance of the DDP calculation during the stress dwell period in a dwell fatigue cycle, the average effective stress (considering both normal and shear components and consistent with the experimental measurement of stress) oscillates within the region of interest (ROI) roughly about the macroscopic applied stress shown Figure 5 (a), which remains constant. We sample the average stress within the ROI every 0.7 s (*i.e.* spanning every 100 DDP time increments), sufficient to resolve the time-dependent dislocation dynamics and hence to capture the stress drops. At least five realisations of the initial dislocation nucleation and obstacle sites were carried out to average across random fluctuations, which also helps collect sufficient stress drop data to perform the statistical analysis in the following sections. Figure 5 (b) shows the dislocation density and average y-component strain against loading time. There is no apparent sudden change in the dislocation density, which indicates that dislocation bursts (such as those observed in micro-pillar compression tests (Benzerga, 2009) and nano-indentation (Balint et al., 2006)) governed by dislocation nucleation events (Agnihotri and Van der Giessen, 2015; Benzerga, 2008; Chakravarthy and Curtin, 2010; Shan et al., 2008; Shishvan and Van der Giessen, 2010) are insignificant within this Ti crystal during stress dwells. Therefore, the rate-sensitivity (*i.e.* time-dependence) of the stress oscillation during the dwell period mainly originates from pinning and unpinning events (Zheng et al., 2016c) where dislocations interact with obstacles. The instantaneous dislocation structures and the yy-stresses at two instants P and Q labelled in Figure 5 (a) are shown in Figure 5 (c-d). Considerable slip takes place that develops by the accumulation of dislocations escaping from obstacles, and hence the local stresses vary with the dwell time as a consequence. Therefore, thermal activation events where dislocations escape obstacles within crystals could potentially be the key mechanism for the stress drops observed experimentally in the HEDM tests.

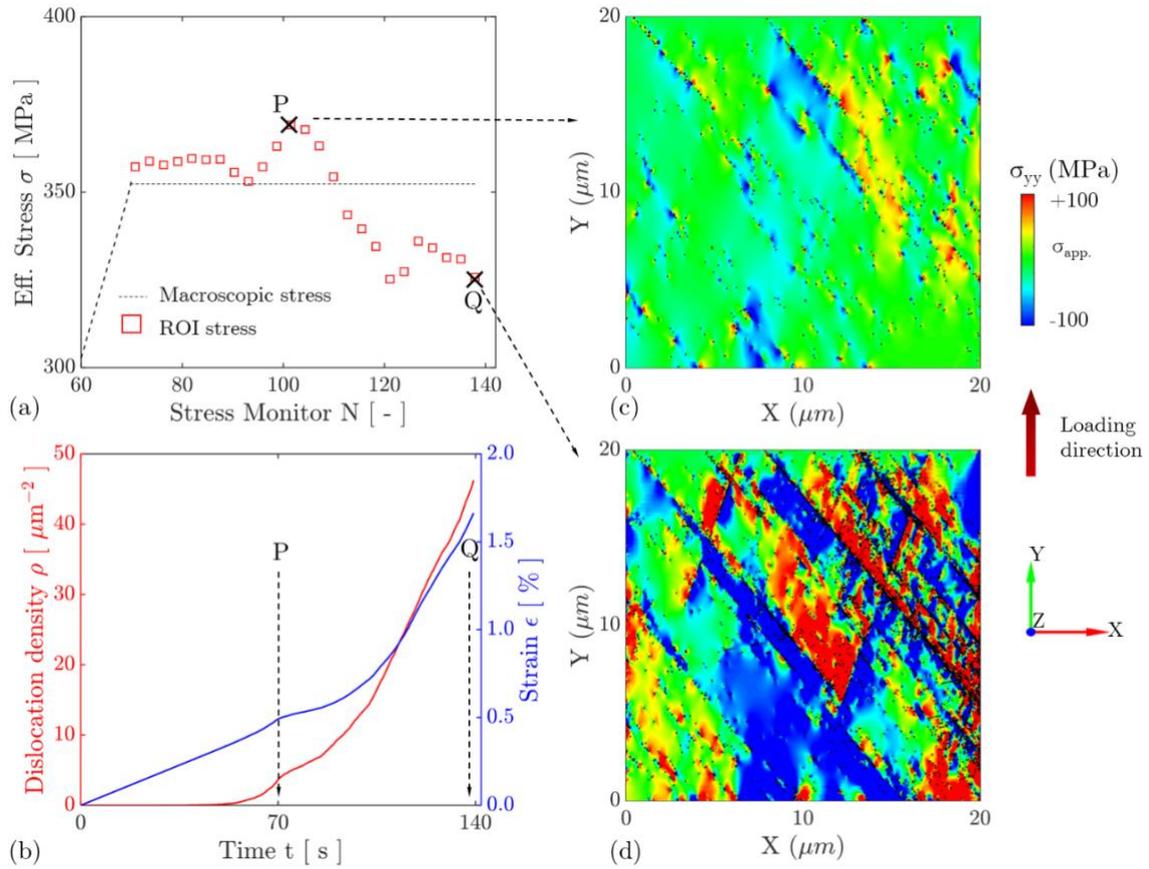

Figure 5. The stress oscillation within a Ti–7Al–O single crystal with the reference crystallographic orientation during a dwell cycle predicted using the DDP method. (a) the macroscopic stress and the average stress in the ROI defined in Figure 4 (d). P and Q denotes the start and end points in time of the stress dwell period. (b) the strain and dislocation density evolution within the ROI. (c) and (d) the stress distribution and instantaneous dislocation structure within the single crystal at instants P and Q, respectively, during a dwell cycle, suggesting significant time-dependent change of dislocation structure and stress due to thermally activated dislocation escape.

Systematic parametric studies using discrete dislocation plasticity have been undertaken in order to investigate the effect of the governing parameters on the stress oscillations of the single crystal shown in Figure 3 and observed in the HEDM creep test shown in Figure 1. Representative thermal activation governing parameters of α-titanium alloys (Xu et al., 2020b; Zheng et al., 2016a), namely the activation energy $\Delta F$ and activation volume $\Delta V$, were chosen, and the stress oscillations are shown in Figure 6 (a) and (b), respectively during the stress hold period. A small value of activation energy and a large value of activation volume both result in high

Page **16** of **32**

amplitude stress oscillations, which has been argued to give rise to the sudden slip events (Crosby et al., 2015; Richeton et al., 2005; Weiss and Marsan, 2003) frequently observed in the loading portion of micro-pillar tests (Dimiduk et al., 2006; Uchic et al., 2004). As the slip events during stress holds arise by dislocation escape from obstacles where they had been pinned (Dunne et al., 2007), the stress oscillations are therefore governed by the thermal activation parameters. The combination of a small activation energy and a large activation volume then gives a large slip rate, e.g. eq.(2). For a given single crystal with a particular combination of thermal activation parameters, crystallographic orientation also affects the stress oscillations through the resolved shear stress (RSS). The dependence of the stress oscillations upon crystallographic orientation is shown in Figure 6 (c), showing a variety of rotations all with their (0002)-axis along Z, the plane strain direction. A crystallographic orientation that facilitates slip (*i.e.* a large macroscopic Schmid factor) generally intensifies the stress drop magnitude, which is consistent with eq.(2). We have investigated the effect of obstacle spacing by altering the obstacle density $\rho_{obs}$, and a lower obstacle density gives rise to larger stress drops. That is, the back stress due to other pinned dislocations inhibits dislocation escape from obstacles, likely to occur in alloy with a large number of obstacles, such as hard participates, interstitial O pinning sites and second phases etc (Zhang et al., 2021). Therefore, the predicted effect of the obstacle spacing reflects the role of oxygen on the stress variation that is revealed in the HEDM tests (see Figure 1).

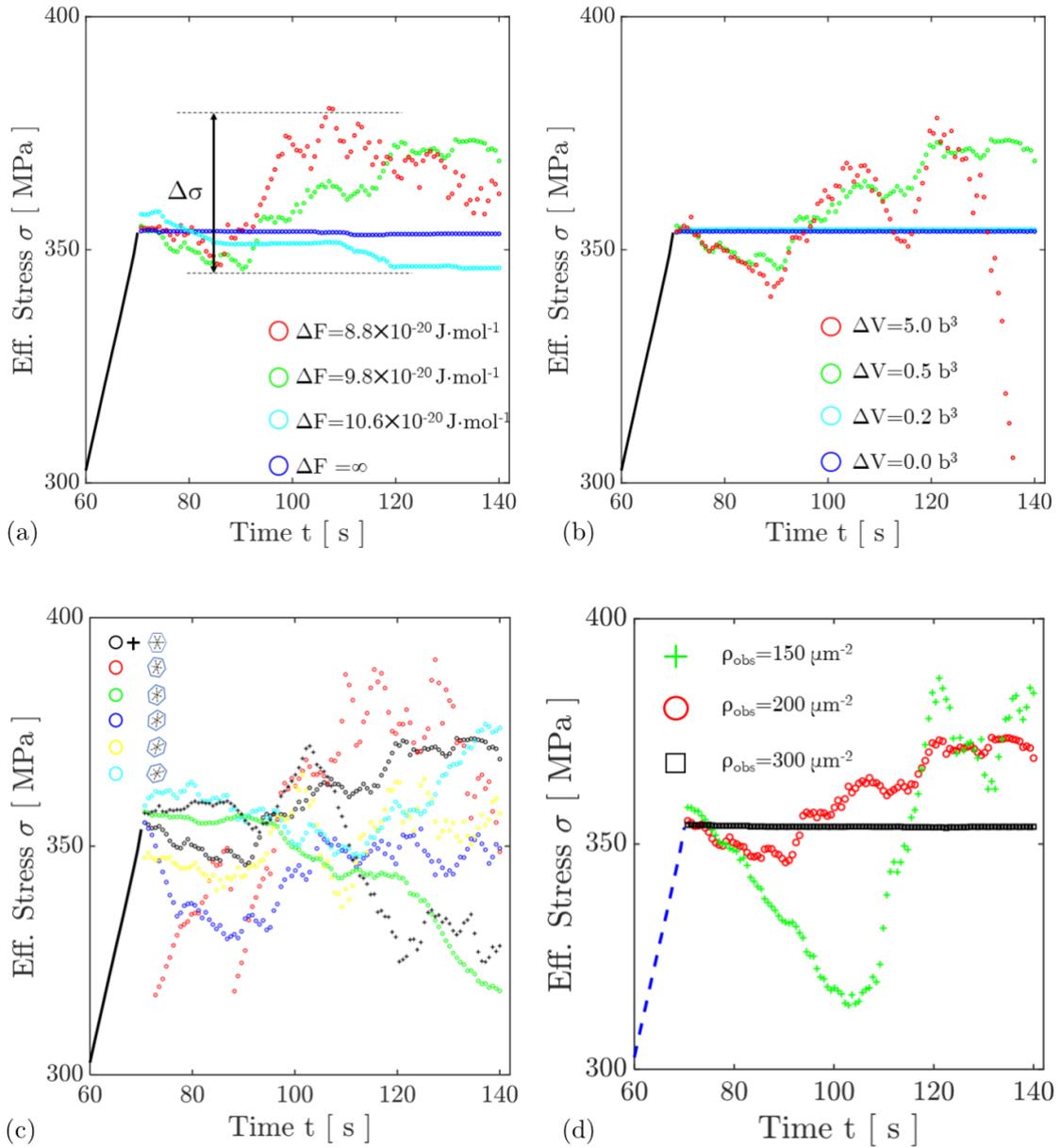

Figure 6. Parametric studies of the stress oscillation in the Ti-7Al-O single crystal during the stress hold period: (a) the activation energy $\Delta F$, (b) the activation volume $\Delta V$, (c) crystallographic orientation (*i.e.* the slip system orientation with respect to the loading direction ranging from 0 to 60 degrees) and (d) the obstacle density $\rho_{obs}$. The strong correlation between the stress oscillations and these governing parameters suggests that the stress drop phenomenon is caused by the dislocation escape from obstacles, which is governed by thermal activation events.

A histogram of the resolved shear stress drops during stress hold is shown in Figure 7(a) for the basal and prism slip systems (α phase, HCP crystal structure), where the data are gathered for a range of crystallographic orientations but with the same materials properties. There are three realisations of the initial random dislocation sources and obstacles for each orientation to fully consider the statistics of dislocation dynamics (Deshpande et al., 2001) and to collect sufficient stress drops. The activation energies of the basal and prism slip systems are taken from our prior independent micro-pillar tests on single crystals of a similar titanium alloy (Zhang et al., 2016), which reports on the intrinsic differences in strain rate sensitivity of these slip systems. Basal slip show statistically larger RSS drops compared to that for prism slip, which is a consequence of the relatively low activation energy. The corresponding complementary cumulative distribution function (CCDF) of normalised shear stress drops shown in Figure 7 (b) suggests the same trend, consistent with the experimental observations in Figure 1(d) and (e). The DDP analysis shows that the stress drops related to α-Ti slip system activation can be explained by dislocation escape events from obstacles driven by thermal activation. In addition, by correctly capturing the differing slip intermittency observations for basal versus prism slip activation, it confirms the rate sensitivity differences between the prism and basal slip systems in α-Ti alloys, which controversially differs from those inferred in some other experimental characterisation work (Xiong et al., 2021). This in turn evidences that cold creep and consequent load shedding taking place in the soft / hard oriented grain combinations is likely to be stronger for basal slip activation in the soft grain as opposed to prism, as reported in (Zhang and Dunne, 2018). This does not preclude the involvement of prism slip in crack initiation in the hard grain, and TEM studies suggest both soft-grain basal and prism slip can lead to hard-grain facet nucleation. This hypothesis has also been previously examined by DDP, *e.g.* (Zheng et al., 2016a) and using crystal plasticity simulations, *e.g.* (Xiong et al., 2021).

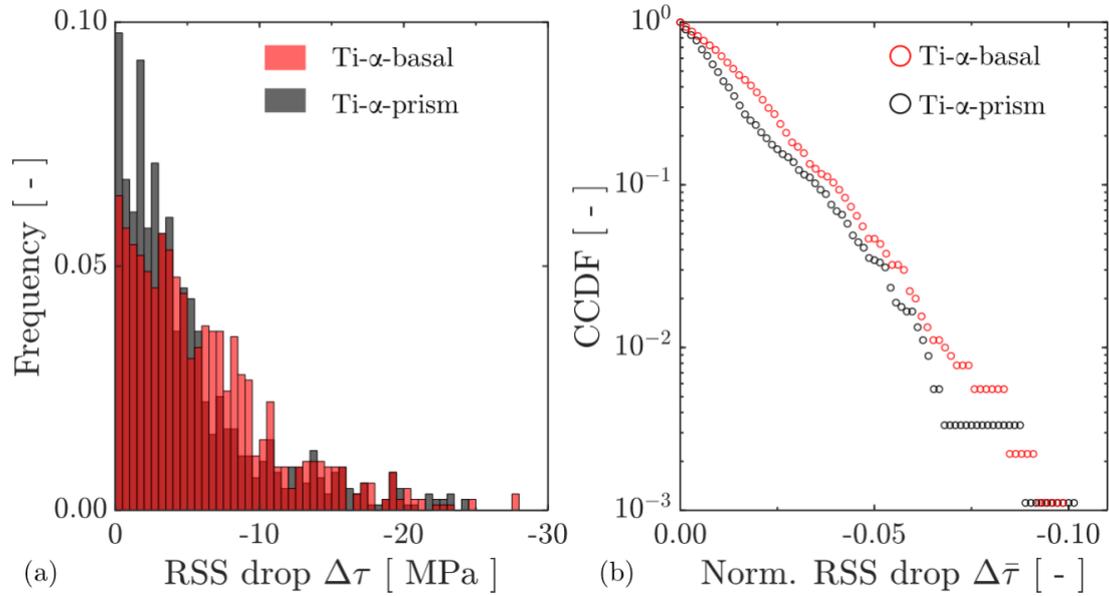

Figure 7. DDP predicted histogram of the resolved shear stress (RSS) drop events and the resulting complementary cumulative distribution functions (CCDF) for normalised resolved shear stress drops on the activated slip systems. (a) and (b) show the behaviours of the basal and prism slip systems within the Ti–7Al–O single crystal with a distribution of crystallographic orientations.

The histogram of the resolved shear stress drops is shown in Figure 8 (a) for single crystals assigned with three different obstacle densities but otherwise with the same material properties. A large value of obstacle density strongly diminishes the stress drops by virtue of hindering further slip while the RSS drops are relatively large in a crystal with sparse obstacles. The corresponding CCDF diagram shown in Figure 8 (b) also reflects the underlying mechanism for stress drops limited by the large number of obstacles, and the slip intermittency diminishes when the obstacle density reaches a sufficiently large value as the thermal activation events are progressively suppressed, Figure 6 (d). The dislocation density evolution and structure shown in Figure 8 (c) also suggests that a low obstacle density (*i.e.* a large obstacle spacing) facilitates thermal activation and slip intermittency. The mechanisms revealed in the numerical investigation of the obstacle spacing using the DDP model elucidate the observed differences between the experimental samples with different levels of oxygen, Figure 1 (d) and (e). The role of the oxygen interstitials is interpreted as deactivating slip intermittency (Barkia et al., 2017; Chong et al., 2020) and stress drops such that there may be a benefit to the dwell fatigue life of Ti alloys due to a reduction in large stress oscillations that might, for example, initiate cracking and slip in adjacent, hard-oriented grains that would otherwise find it difficult to activate plastic deformation.

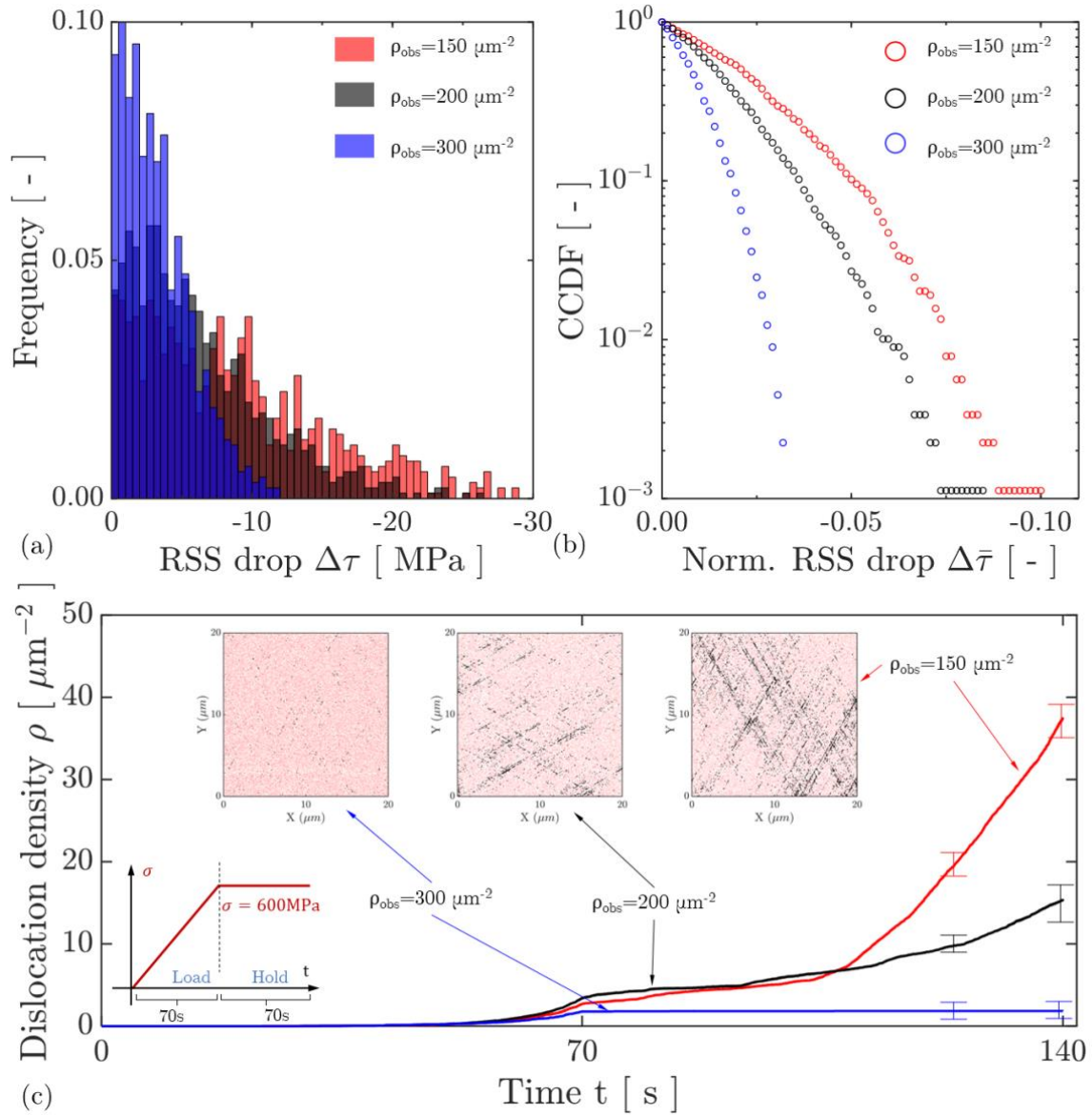

Figure 8. The effect of dislocation obstacle density on the stress drop events. (a) and (b) show the DDP-predicted histogram of the resolved shear stress (RSS) drop events and the resulting complementary cumulative distribution functions (CCDF) for normalised resolved shear stress drops on the activated slip systems with three different obstacle densities. (c) the dislocation structure (inset figures) and density evolution (line plot) within the crystals. The black and red dots represent dislocations and obstacle sites, respectively, in the inset figures. The error bars along the dislocation density curves are collected based on repeated simulations on a number of realisations, reflecting the randomness of the dislocation dynamics.

The dislocation structures observed by TEM in the Ti–Al–O samples subject to stress relaxation with different O contents were shown in Figure 2. The corresponding loading history (*i.e.* a 130 s strain-controlled load-up period to 1.5% macroscopic strain followed by a 600 s strain-hold period) has been applied to a DDP model of a Ti–Al–O single crystal with three obstacle densities. The crystal orientation is configured with $[01\bar{1}1]$ close to the loading direction, which resembles the stress relaxation test setup. The dislocation density evolution and dislocation structure during the stress relaxation test predicted using the DDP model and the calibrated material properties are shown in Figure 9. In contrast to the creep simulation, a larger obstacle density yields a higher dislocation density response, which is consistent with the expected hardening effect of obstacles (Chakravarthy and Curtin, 2010) through dislocations piling up under strain-controlled loading (Deshpande et al., 2005). The instantaneous dislocation structures with the corresponding obstacle distribution in the single crystal with three obstacle densities are shown in the inset figures, and the results are extracted at the end of the stress relaxation. A considerably larger number of dislocations are observed (roughly twice as many as in the low-obstacle-density grain) in the <a> prism planes in the grain with a high obstacle density. As O atoms can be considered as obstacles for dislocation gliding in titanium alloys, the DDP simulation is shown to reflect the obstacle density effect that has been shown by TEM observations, *i.e.* that high oxygen levels lead to a high dislocation density during stress relaxation tests, Figure 2.

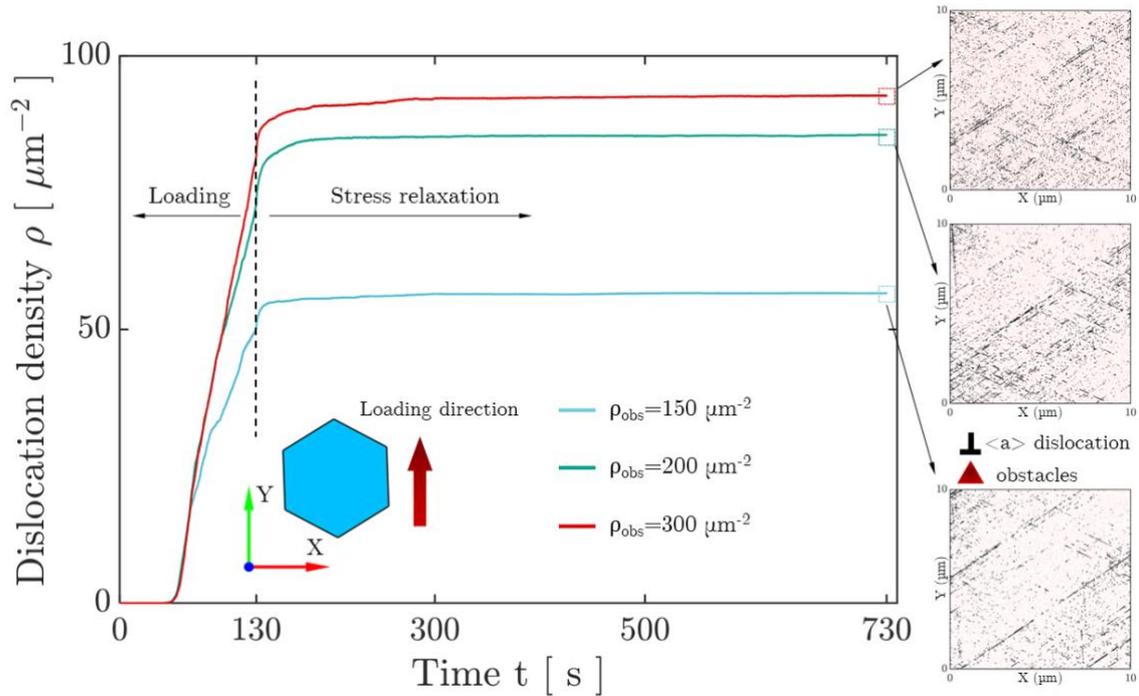

Figure 9. Effect of obstacle density on the dislocation density and dislocation structures during a stress relaxation loading cycle. The dislocation density is averaged within the whole single crystal, with orientation $[01\bar{1}1]$ in the loading direction, as depicted in the inset figure. Three typical obstacle densities are assigned to the single crystal, which could represent e.g. various levels of interstitial oxygen content. The dislocation structures are shown next to the line plot for the single crystals at the end of the stress relaxation.

The representative time for thermally activated escape $t_{obs}$ that describes the duration of dislocations pinning at obstacles is shown in Figure 10 for a range of common near-α, α+β and more heavily β-stabilised Ti alloys at room temperature. The escape times are calculated using eq.(2), and the range of resolved shear stresses (*i.e.* not the applied stress) exerted on individual dislocations is chosen to cover representative values observed in experiments, *e.g.* (Kishida et al., 2020; Xu et al., 2020a) and discrete dislocation calculations (Zheng et al., 2016a). The governing parameters, namely activation volume $\Delta V$ and activation energy $\Delta H$, are obtained from our previous calibration experiments (see Section 3.2) and from the literature (Xu et al., 2020b; Zhang et al., 2015; Zheng et al., 2016a; Zheng et al., 2016b). A short escape time statistically indicates that an alloy is more rate-sensitive at a given stress level as dislocations tend to be released from individual obstacles more easily, which is followed by potential slip intermittency. In addition, dwell fatigue is more detrimental to the Ti alloys with the escape time matching the typical dwell period, which is indicated by the blue shaded region (1-3 minutes). Considering a range the alloy properties (including those considered in this study), the alloy Ti6246 shows a relatively low rate-sensitivity at room temperature (Qiu et al., 2014; Zheng et al., 2017) which tends to give a minimal effect of dwell fatigue, as its escape time considerably exceeds the dwell time period for low stresses. However, alloys Ti-6242 and Ti-7Al are indicated to be more sensitive to dwell fatigue as their escape times are well aligned with the dwell period. Therefore, the escape time of dislocations pinned at obstacles, which is governed by the thermal activation of dislocation escape, has been shown to be strongly correlated to the slip intermittency during stress dwell, and can be used as an indicator for evaluating the dwell performance of the considered Ti alloys. This understanding can help guide microstructure optimisation for design and development of new Ti alloys.

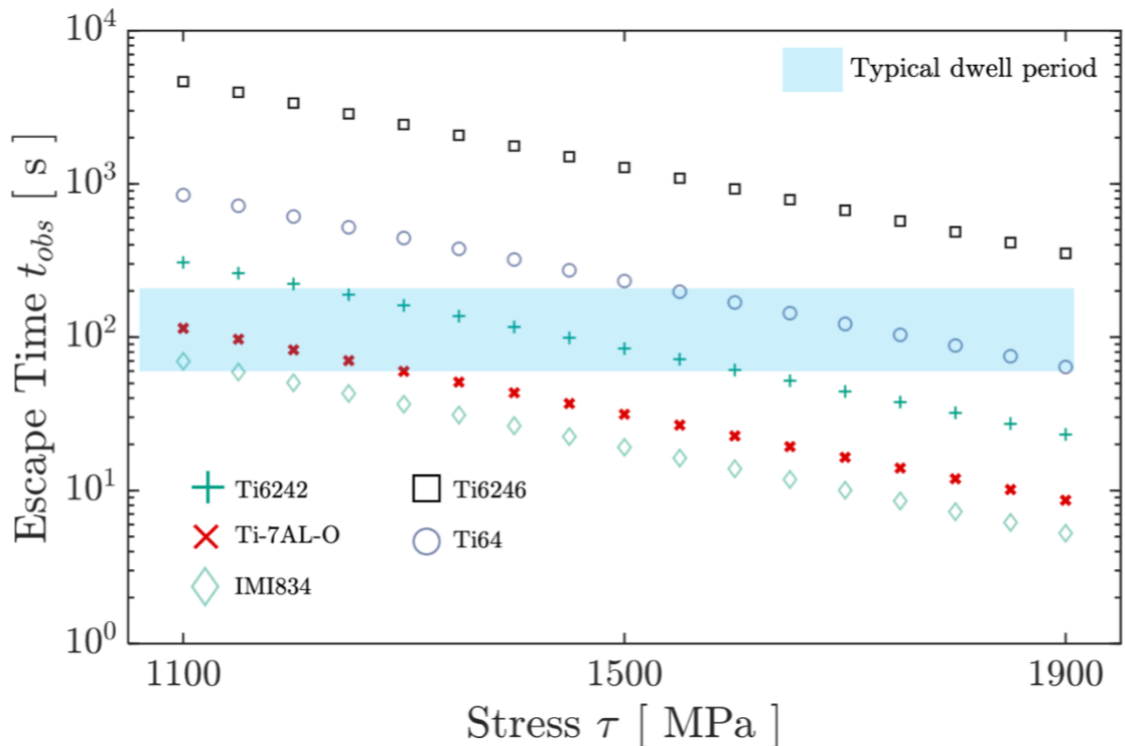

Figure 10. The analytical escape time of dislocations pinned at obstacles for various representative Ti alloys subject to a range of shear stresses exerted on individual dislocations. The escape time is calculated using eq.(2) that considers different combinations of activation energy and activation volume of the Ti alloys investigated. The blue band indicates the representative dwell period in a cycle.

## 5 Conclusions

Discrete dislocation plasticity analysis has been performed to investigate the slip intermittency and stress oscillations experimentally observed previously within Ti–7Al–O alloys during stress holds in dwell fatigue cycles. The underpinning mechanisms for the sudden slip intermittency and stress oscillations are shown to be consistent with dislocations escaping from pinning obstacles through thermal activation.

- The events of slip intermittency and stress oscillation are governed by the thermal activation governing parameters, crystallographic orientation and obstacle spacing.
- Slip along <a>-basal systems show stronger stress oscillations than prismatic ones, using the thermal activation properties previously obtained from single crystal pillar compression tests, which is statistically consistent with our recent high energy X-ray diffraction microscopy (HEDM) observations.

- Samples with lower obstacle density show larger stress drops, which provides mechanistic understanding for the role of interstitial oxygen in slip intermittency and the resulting dislocation structures, and hence helps guide the design of Ti alloys subject to dwell fatigue.

## Acknowledgements

YX and FPED acknowledge the financial support by the Engineering and Physical Sciences Research Council for funding through the grant EP/R018863/1, and DD through EP/t01041X/1. FFW was funded by Rolls-Royce plc and by the EPSRC Centre for Doctoral Training in the Advanced Characterisation of Materials (EP/L015277/1). Much of this work was initiated and funded through the Hexmat EPSRC program grant EP/K034332/1. DD also acknowledges provision of a Royal Society Industry Fellowship.